\documentstyle[12pt]{article}
\newlength{\bredde}
\def\slash#1{\settowidth{\bredde}{$#1$}\ifmmode\,\raisebox{.15ex}{/}
\hspace*{-\bredde} #1\else$\,\raisebox{.15ex}{/}\hspace*{-\bredde} #1$\fi}
\newcommand{\beq}{\begin{equation}}
\newcommand{\eeq}{\end{equation}}
\def\beqn{\begin{eqnarray}}
\def\eeqn{\end{eqnarray}}
\def\sepand{\rule{14cm}{0pt}\and}
\def\gtwid{\raise.3ex\hbox{$>$\kern-.75em\lower1ex\hbox{$\sim$}}}
\def\ltwid{\raise.3ex\hbox{$<$\kern-.75em\lower1ex\hbox{$\sim$}}}
\begin{document}
\topmargin -0.8cm
\oddsidemargin -0.8cm
\evensidemargin -0.8cm
\title{\Large{
Generalized Lagrangian Master Equations}}

\vspace{0.5cm}

\author{{\sc Jorge Alfaro} \\
{Fac. de Fisica} \\ {Universidad Catolica de Chile}\\
{Casilla 306, Santiago 22, Chile} \\
\sepand
{\sc Poul H. Damgaard} \\
CERN -- Geneva, Switzerland}
\maketitle
\vfill
\begin{abstract} We discuss the geometry of the Lagrangian quantization
scheme based on (generalized) Schwinger-Dyson BRST symmetries. When a
certain set of ghost fields are integrated out of the path integral,
we recover the Batalin-Vilkovisky formalism, now extended to arbitrary
functional measures for the classical fields. Keeping the ghosts
reveals the crucial role played by a natural connection on the space of fields.
\end{abstract}
\vfill
\begin{flushleft}
CERN--TH-7247/94 \\
May 1994 \\
hep-th/9405112
\end{flushleft}
\newpage

\setcounter{page}{1}

The Lagrangian quantization scheme of Batalin and Vilkovisky
\cite{Batalin} has a direct relation to what we have called
``Schwinger-Dyson BRST symmetry", -- the BRST symmetry whose
Ward identities provide the most general Schwinger-Dyson equations
of any given quantum theory \cite{us0,us1}. Imposing this
Schwinger-Dyson symmetry on the theory leads immediately to
a Lagrangian Master Equation \cite{us}, which reduces to the
Batalin-Vilkovisky Master Equation \cite{Batalin} upon
integrating out a certain set of new ghost fields $c^A$. The
``antifields" of the Batalin-Vilkovisky formalism are nothing
but the usual antighosts of these new fields $c^A$ \cite{us}.

The easiest way to see the need for new ghost fields $c^A$ is
to derive the Schwinger-Dyson BRST symmetry from a particular
collective field formalism \cite{us1}. Since the BRST symmetry
in question is related to arbitrary local shifts of all field
variables, there is a one-to-one correspondence between all
fundamental fields $\phi^A$ of a given theory and the required
collective fields $\varphi^A$. The appearance of both the new
ghosts $c^A$ and the collective fields $\varphi^A$ is not
fortuitous. For example, if one wishes to quantize a theory
in such a manner that it is invariant under BRST and anti-BRST
symmetry simultaneously \cite{Ext}, then both of these new fields
can simply not be removed from the Master Equation \cite{Ext1}.
The new ghosts $c^A$ also play an important r\^{o}le when one
derives the Lagrangian BRST quantization from the BFV theorem
of the Hamiltonian formalism \cite{Frank}.

Gauge field theories can be dealt with at the same level as
theories without internal gauge symmetries. The solution to the
quantization problem is then entirely given by imposing the
Schwinger-Dyson BRST symmetry, and demanding certain
boundary conditions on the resulting differential equation.
Information about the internal gauge symmetries enters only at the
stage where boundary conditions are imposed. These boundary
conditions can be chosen to equal those of ref. \cite{Batalin},
but more general procedures are also possible \cite{us}.

The Schwinger-Dyson BRST symmetry is intimately related to the
BRST symmetry of field redefinitions \cite{us2}. This is not
surprising, because Schwinger-Dyson equations can be viewed as
the tool with which to describe the quantized theory
independently of a specific path integral representation. In
fact, the Schwinger-Dyson BRST symmetry is precisely the
gauge-fixed remnant of a hidden local gauge symmetry present in
any quantum field theory: The gauge symmetry of local field
reparametrizations \cite{us1}. Ordinarily one chooses from the
outset a basis of field variables with which to describe physics,
but the field redefinition theorem ensures  --  at least under
certain mild assumptions about the asymptotic states -- that any
other choice of variables should describe the same physics.
Technically, this can be seen from the invariance of S-matrix
elements under field redefinitions. Invariance of the S-matrix
under such reparametrizations is precisely a reflection of the
local gauge symmetry of field redefinitions \cite{us1}, in just
the same manner as invariance of S-matrix elements under
internal gauge transformations reflects the ordinary gauge
symmetry of gauge field theories.

Since the Schwinger-Dyson BRST symmetry can be viewed as one
particular facet of the general field reparametrization BRST
symmetry, one would expect that a more general Lagrangian
quantization scheme could be derived from the latter. This should
provide a quantization principle independent of the field
representation, ``covariant" in the space of field variables.
Such a generalized quantization procedure should by definition
be closely related to the geometric formulation of the
Batalin-Vilkovisky formalism, a subject that has recently received
considerable attention \cite{Geometry,Schwarz}.

The aim of the present paper is to derive this more general
covariant Lagrangian quantization prescription starting from
the generalized Schwinger-Dyson BRST symmetry, -- the BRST symmetry
of field redefinitions. In the process we hope to add some
physical insight to the more abstract algebraic considerations
of refs. \cite{Geometry,Schwarz}. Our manipulations will throughout
be ``formal" in the sense that we shall employ standard manipulations
in the path integral, assuming the existence of a suitable
symmetry-preserving regulator. Some of the subtleties involved
in this process, especially at the two-loop level, are discussed
in ref. \cite{us1}.

To set the stage for the generalizations that are to follow, let
us first briefly consider the simplest case, that of an action free
of internal gauge symmetries. Fields of the classical action $S$
are denoted by $\phi^A$; they can be of arbitrary Grassmann parity
$\epsilon(\phi^A) \equiv \epsilon_A$.\footnote{Our conventions are
described in detail in the appendix of ref. \cite{us}.}
The index $A$ labels collectively all internal quantum numbers and
space-time variables. A quantum action $S_{ext}$ that incorporates
the correct Schwinger-Dyson BRST symmetry can in this case be taken
to be simply \cite{us0,us1}: $
S_{ext}[\phi,\phi^*,c] = S[\phi] - \phi^*_Ac^A$,
with a new ghost-antighost pair $c^A, \phi^*_A$ of Grassmann parities
$\epsilon(c^A) = \epsilon(\phi^*_A) = \epsilon_A + 1$. Their ghost
number assignments are $gh(c^A) = - gh(\phi^*_A) = 1$. With this set
of fields, the Schwinger-Dyson BRST symmetry reads
\beq
\delta \phi^A = c^A~,~~~
\delta c^A = 0~,~~~
\delta \phi^*_A = - \frac{\delta^l S}{\delta\phi^A} ~.
\eeq
In this simple case, it is obviously possible to substitute $S_{ext}$
for $S$ in the transformation law for $\phi^*_A$, but in general
care is required in such a substitution. By correct Schwinger-Dyson
equations, we shall always refer to those that formally follow for
the classical fields of the classical action, independently of
whether the path integral has been given a precise meaning through
an appropriate gauge fixing, when needed.

The above choice incorporates the Schwinger-Dyson BRST symmetry (1)
in the particular field variables $\phi^A$. To find a more covariant
formulation, let us perform a field redefinition of all the classical
fields $\phi^A$. At this stage we restrict ourselves to redefinitions
that do not mix in the new ghost fields $c^A, \phi^*_A$. We follow
to a large extent the formulation presented in ref. \cite{us2}.
Denote the new field variables by $\Phi^A$, and the transformation
by $F$. Introduce left ($L$) and right ($R$) vielbeins $e^A_{(L,R)B}$ and
their inverses, $E^A_{(L,R)B}$,
through the definition
\beq
e^A_{(L,R)B}(\Phi) \equiv \frac{\delta^{l,r} F^A(\Phi)}{\delta \Phi^B}~,~~~~~
e^A_{(L)B}E^C_{(L)A} =  e^C_{(R)A}E^A_{(R)B} = \delta^C_B ~.
\eeq
We next choose to let the ghost-antighost pair transform
oppositely under $F$, $i.e.$, in total:
\beq
\phi^A = F^A(\Phi)~,~~~
C^A = E^A_{(R)B}c^B~,~~~
\Phi^*_A = \phi^*_Be^B_{(R)A} ~,
\eeq
where $C^A$ and $\Phi^*_A$ are the new transformed ghost fields.
This has the advantage that the ghost-antighost measure formally, or
with a suitable symmetry-respecting regulator, remains invariant
under the transformation.
Of course, the $\phi^A$-measure will in general {\em not} remain
invariant, but acquire a Jacobian factor $\sqrt{g}$, where $g$
is the superdeterminant of the metric
\beq
g_{AB}(\Phi) = \eta_{CD}e^C_{(L)A}(\Phi)e^D_{(R)B}(\Phi) ~.
\eeq

Consider now the action $S_{ext}$. Since it must transform as a scalar
under $F$, we immediately have, using (3),
\beq
S_{ext} = S[F(\Phi)] - \Phi^*_AC^A ~.
\eeq
This transformed action is invariant under the transformed
Schwinger-Dyson BRST symmetry
\begin{eqnarray}
\delta \Phi^A = C^A~,~~~
\delta C^A = 0~,~~~
\delta \Phi^*_A = (-1)^{\epsilon_M+1}\Gamma^M_{AK}C^K\Phi^*_M -
\frac{\delta^l S}{\delta \Phi^A} ~.
\end{eqnarray}
It is also straightforward to check that the functional measure is
formally invariant. The ``connection" $\Gamma^A_{BC}$ is the
(superspace) Christoffel symbol of second kind
\cite{Arnowitt},
\beq
\Gamma^A_{BC} \equiv \frac{1}{2}(-1)^{\epsilon_A\epsilon_C}
\left[(-1)^{\epsilon_C\epsilon_D}\frac{\delta^r
g_{BD}(\Phi)}{\delta\Phi^C} + (-1)^{\epsilon_B+\epsilon_C+
\epsilon_B\epsilon_C+\epsilon_B\epsilon_D}\frac{\delta^r g_{CD}(\Phi)}
{\delta\Phi^B} - \frac{\delta^r g_{BC}(\Phi)}{\delta \Phi^D}\right]
g^{DA}~.
\eeq

In the new set of coordinates, the Schwinger-Dyson equations are
Ward identities $0 = \langle \delta[\Phi^*_A G(\Phi)] \rangle$. In
detail, with $\Gamma^M_{AM} = (-1)^{\epsilon_M}(\sqrt{g})^{-1}\delta^l
(\sqrt{g})/\delta\Phi^A$,
\begin{eqnarray}
0 = \frac{i}{\hbar}(-1)^{\epsilon_G+1}\langle \delta[\Phi^*_A G(\Phi)] \rangle
= \left\langle (-1)^{\epsilon_M}\Gamma^M_{AM}(\Phi)G(\Phi) +
\left(\frac{i}{\hbar}\right)\frac{\delta^l S}
{\delta \Phi^A}G(\Phi) + \frac{\delta^l G}{\delta \Phi^A} \right\rangle ~,
\end{eqnarray}
where in the last bracket we have integrated out the ghost-antighost
pair in order to compare with the conventional formulation of
field-covariant Schwinger-Dyson equations. Such equations
are normally derived from the invariance of the measure $[d\Phi]$
under arbitrary local shifts, $i.e.$, from
\beq
0 = Z^{-1}\int [d\Phi]\sqrt{g(\Phi)}\left[(\sqrt{g(\Phi)})^{-1}
\frac{\delta^l}{\delta\Phi^A}\left\{e^{-S[\Phi]}\sqrt{g(\Phi)}
G(\Phi)\right\}\right]~.
\eeq
In contrast, in the present formulation these equations are
automatically incorporated into the action principle.

The Master Equation for the action $S_{ext}$ in transformed
coordinates is derived in as trivial a manner as in the original
variables; it is simply the statement that $S_{ext}$ is invariant
under the BRST symmetry (6). Thus $0 = \delta S_{ext}$ immediately
gives
\beq
\frac{\delta^r S_{ext}}{\delta \Phi^A}C^A =
\frac{\delta^r S_{ext}}{\delta \Phi^*_A}\frac{\delta^l S}{\delta \Phi^A}
= \frac{\delta^r S_{ext}}{\delta \Phi^*_A}\frac{\delta^l S_{ext}}
{\delta \Phi^A}~.
\eeq
The extra term in the transformation law for $\Phi^*_A$
in eq. (6), which is proportional to the connection $\Gamma^A_{BC}$,
does not contribute to the Master Equation due to the symmetry
properties of $\Gamma^A_{BC}$ and the ghosts $C^A$.
The Master Equation (10) is of precisely the same form as that of the
original $S_{ext}$ \cite{us}, except that it is now expressed in the
new coordinates.

To extend this construction to field theories in all generality,
including those of arbitrarily complicated gauge-symmetry structure,
one can proceed by demanding that the above coordinate-covariant
Schwinger-Dyson equations for the classical fields are satisfied
at the formal level throughout, and even before any gauge
fixings. A sufficient, but perhaps not necessary, condition is that
the ghosts $C^A$ enter only linearly, and only in the combination
$\Phi^*_AC^A$, as in eq. (5). This ensures that the crucial
integral over $C^A$ and $\Phi^*_A$ is diagonal, and in particular
that $\langle C^A\Phi^*_B \rangle = -i\hbar\delta^A_B$, an
ingredient needed in eq. (8) to recover the correct
Schwinger-Dyson equations. In general, on should not expect to be
able to split the extended action $S_{ext}$ into the form
$S_{ext}[\Phi,\Phi^*,C] = S[\Phi] - \Phi^*_AC^A$, as in eq. (5).
But the above requirement is equivalent to demanding that $S_{ext}$
is of the form $S_{ext}[\Phi,\Phi^*,C] = S^{BV}[\Phi,\Phi^*]
- \Phi^*_AC^A$, where $S^{BV}$ is simply everything left over after
the term linear in $C^A$ has been taken out.

Correct Schwinger-Dyson equations are obtained even if
the extended action $S_{ext}$ is not invariant under the
Schwinger-Dyson BRST symmetry, but only transforms in precisely
such a manner as to cancel a perhaps non-trivial Jacobian factor
from the functional measure.
In the {\em old} coordinates, an arbitrary action $S_{ext}[\phi,
\phi^*,c] = S^{BV}[\phi,\phi^*] - \phi^*_Ac^A$ which satisfies the
full quantum Master Equation \cite{us}
\beq
\frac{1}{2}(S_{ext},S_{ext}) = - \frac{\delta^r S_{ext}}{\delta
\phi^A} c^A + i\hbar \Delta S_{ext} ~,
\eeq
gives rise to the Batalin-Vilkovisky Master Equation \cite{Batalin}
for $S^{BV}$:
\beq
\frac{1}{2}(S^{BV},S^{BV}) = i\hbar \Delta S^{BV}~.
\eeq
Here $(\cdot,\cdot)$ is the antibracket, and $
\Delta \equiv (-1)^{\epsilon_A+1}\frac{\delta^r}{\delta \phi^A}
\frac{\delta^r}{\delta \phi^*_A}$
is the correction term from the measure \cite{Batalin}. Both can
straightforwardly be derived from the Schwinger-Dyson BRST
symmetry \cite{us}.

To find the generalized Master Equation in the {\em new} coordinates,
we must be careful when expressing the BRST transformation laws of
all fields, ghosts and antighosts in terms of the new variables only.
In particular, since $\Phi^*$ in general will enter non-trivially
apart from the term $\Phi^*_AC^A$, and since these antighosts
arise from a $\Phi$-dependent transformation, a new implicit
$\Phi$-dependence enters through $\Phi^*$:
\beqn
\delta \Phi^A &=& C^A \cr
\delta C^A &=& 0\cr
\delta \Phi^*_A &=& (-1)^{\epsilon_M+1}\Gamma^M_{AK}C^K\Phi^*_M +
(-1)^{\epsilon_A\epsilon_M+1}\frac{\delta^r S^{BV}}{\delta\Phi^*_K}
\Gamma^M_{KA}\Phi^*_M - \frac{\delta^l S^{BV}}{\delta\Phi^A} ~.
\eeqn
With these transformation rules it is easy to get the following
master equation:
\beqn
\frac{\delta^r S^{BV}}{\delta \Phi^*_A}\frac{\delta^l S^{BV}}
{\delta\Phi^A} = i\hbar (-1)^{\epsilon_A}
\frac{1}{\sqrt{g}}\frac{\delta}{\delta \Phi^A}\left(\sqrt{g}
\frac{\delta S^{BV}}{\delta\Phi^*_A}\right) ~.
\eeqn

The operator
\beq
\Delta_\rho \equiv (-1)^{\epsilon_A+1}
\frac{1}{\sqrt{g}}\frac{\delta^r}{\delta \Phi^A}
\left(\sqrt{g}\frac{\delta^r}{\delta \Phi^*_A}\right) ~,
\eeq
associated with the measure density $\rho = \sqrt{g}$,
is the covariant generalization of the Batalin-Vilkovisky operator
$\Delta$ of eq. (12). Its form can also be inferred from
general covariance arguments \cite{Geometry,Schwarz}. Here, it arises
straightforwardly from the non-trivial Jacobian factor associated
with the BRST transformation (13). Since we have so far restricted
ourselves to field transformations among the $\phi$'s only, the
resulting measure density $\rho$ does not depend on $\Phi^*$.

In the case of flat coordinates, there is an interesting
direct relation between
the Schwinger-Dyson BRST operator (1) and the operator $\Delta$
\cite{us}. Namely, if one integrates out the ghosts $c^A$ but keeps
the antighosts $\phi^*_A$ in the path integral, the operator
$\Delta$ appears as a ``quantum deformation" (proportional to
$\hbar$) of the BRST operator $\delta$ left over when integrating
out the $c^A$-fields. The quantum deformation of the BRST operator
in the conventional Batalin-Vilkovisky formalism has been
discussed in ref. \cite{Henneaux}. It must be emphasized that the
appearance of this quantum deformation in the BRST operator is
completely unrelated to the appearance of possible quantum
corrections in the Lagrangian Master Equation (12). The quantum
correction in the Schwinger-Dyson BRST operator in the unusual
form in which
the ghosts $c^A$ (but not their antighosts $\phi^*_A$) have been
integrated out of the functional integral is {\em always} present.
The quantum correction to the Master Equation (12) is non-vanishing
only in those particular cases where the functional measure is not
invariant under the Schwinger-Dyson BRST symmetry (independently of
whether the ghosts $c^A$ have been integrated out or not).

In the covariant case we have seen that $\Delta$ in the Master
Equation is replaced by the covariant $\Delta_{\rho}$. Let us
now consider integrating out the new ghosts $C^A$ from the path
integral, and trace what happens to the Schwinger-Dyson BRST
operator in this process. As in the flat case \cite{us}, the
simple identity
\beq
\int [dC] F(C^B)\exp\left[-\frac{i}{\hbar}\Phi^*_AC^A\right]
= F\left(i\hbar\frac{\delta^l}{\delta\Phi^*_B}\right)
\int [dC]\exp\left[-\frac{i}{\hbar}\Phi^*_AC^A\right]
\eeq
is useful here. Consider, inside the path integral,
the BRST variation of an arbitrary functional $G[\Phi,\Phi^*]$.
Using (16) above, we get
\begin{eqnarray}
&& \!\!\!\!\!\!\!\!\!\!\!\!\!\!\!\!\!\!\!\!\!\!\!\delta
G[\Phi,\Phi^*] \cr &=& \frac{\delta^r G}{\delta\Phi^A}C^A
+ \frac{\delta^r G}{\delta\Phi^*_A}\left\{(-1)^{\epsilon_M+1}\Gamma^M_{AK}C^K
\Phi^*_M +
(-1)^{\epsilon_A\epsilon_M+1}\frac{\delta^r S^{BV}}{\delta\Phi^*_K}
\Gamma^M_{KA}\Phi^*_M - \frac{\delta^l S^{BV}}{\delta\Phi^A}\right\} \cr
&\to& \frac{\delta^r G}{\delta \Phi^A}\frac{\delta^l S^{BV}}
{\delta \Phi^*_A} - \frac{\delta^r G}{\delta \Phi^*_A}
\frac{\delta^l S^{BV}}{\delta \Phi^A} +
(i\hbar)\left((-1)^{\epsilon_A}\frac{\delta^r}{\delta \Phi^A}
+ (-1)^{\epsilon_A\epsilon_G+\epsilon_M}\Gamma^M_{AM}\right)
\frac{\delta^r}{\delta \Phi^*_A} G~,
\end{eqnarray}
where the arrow indicates that partial integrations are required
inside the functional integral. Since
\beq
\Delta_{\rho}G = (-1)^{\epsilon_A+1}\frac{1}{\sqrt{g}}\frac{\delta^r}
{\delta \Phi^A} \left(\sqrt{g}\frac{\delta^r G}{\delta \Phi^*_A}\right) =
\left((-1)^{\epsilon_A+1}\frac{\delta^r}{\delta \Phi^A} +
(-1)^{\epsilon_A\epsilon_G+\epsilon_M+1}\Gamma^M_{AM}\right)
\frac{\delta^r}{\delta \Phi^*_A}G~,
\eeq
the equivalent of the Schwinger-Dyson BRST operator after having
integrated out the ghosts $C^A$ is indeed, as expected, given by
\beq
\delta = (\,\cdot\,,S^{BV}) - i\hbar \Delta_{\rho}
\eeq
in the covariant formulation.\footnote{Note that the operator
$\Delta_\rho$ is nilpotent for any $\sqrt{g}$
that depends only on $\phi^A$.} This form of the ``quantum BRST
operator" in the covariant Batalin-Vilkovisky formulation was first
considered by Hata and Zwiebach \cite{Geometry}. Here we see that
it can be derived straightforwardly from the Schwinger-Dyson BRST
operator by integrating out the ghosts $C^A$. It is only because
one chooses such an asymmetric procedure as that of integrating out the
ghosts, while keeping the antighosts in the path integral, that one
has to face the unusual situation of having a quantum correction to
the BRST operator. The {\em full} Schwinger-Dyson operator (13), with
the ghosts $C^A$ kept, automatically includes both classical and
quantum parts, as is customary in quantum field theory.

So far everything has been derived from the flat case using a general
coordinate transformation. In effect, all this amounts to is a
formulation of the Lagrangian BRST quantization scheme in arbitrary
curvilinear coordinates. It is worthwhile to first look at the
quantization problem from the point of view of having been given a
``space of fields'' on which the path integral is to be defined. What
is the effect of {\em curvature} in such a
space of fields? To see a possible consequence, we need to go
back and determine the Schwinger-Dyson equations on such spaces.
As we have seen, once we have the correct Schwinger-Dyson BRST algebra,
the quantization prescription follows immediately.

The correct Schwinger-Dyson equations for field theories defined on
field spaces with a non-vanishing Riemann tensor (but with zero
torsion, see below) can be derived as soon as the functional
integral on such spaces is decided upon. Taking it to be of the form of
a scalar density function $\rho(\phi) = \sqrt{g(\phi)}$,
it is obvious that the Schwinger-Dyson equations (and the
Schwinger-Dyson BRST algebra (13) that reproduces them) are of exactly
the same kind as in eq. (8). This means that the whole quantization
procedure, the Lagrangian Master Equation (14) and the form of the BRST
operator (13), carry over directly to this case without
modifications.\footnote{
The only non-trivial aspect lies in the choice of appropriate boundary
conditions for the Master Equation. In contrast to the simple curvilinear
case, we may not simply take the standard Batalin-Vilkovisky boundary
conditions for Cartesian coordinates and then perform the required
field redefinition to obtain the corresponding boundary conditions in
new coordinates.}
Whereas the case of curvature in the space of fields can thus be treated
straightforwardly, a non-trivial aspect enters if we consider
field spaces with torsion. We shall return to a discussion of this
point elsewhere.

We shall now approach the quantization problem from a different point of
view. Suppose we are {\em given} a measure density $\rho(\phi)$, and the set
of transformations that leave the functional measure $d\phi\rho(\phi)$,
but not the action $S[\phi]$,
invariant. We denote these transformations by
\beq
\phi^A(x) = g^A(\phi'(x), a(x))~,
\eeq
where $a^i(x)$ is a local field parametrizing the transformations. We choose
coordinates such that $g^A$ reduces to the identity at $a^i(x) = 0$.
Invariance
of the functional measure implies a set of identities, generalized
Schwinger-Dyson equations:
\beq
\left\langle\left.\frac{\delta^l g^A}{\delta a^i}\right|_{a=0}
\left[\frac{\delta^l F}{\delta \phi^A} + \frac{i}{\hbar}\frac{\delta^l S}
{\delta \phi^A}F[\phi]\right]\right\rangle = 0~,
\eeq

These Schwinger-Dyson equations are different in form from those obtained
by exploring invariance of the measure $d\phi$ under local shifts. But
under the conditions stipulated below they
have the same content, and can, in fact, be mapped onto one
another. To regain the usual Schwinger-Dyson equations from the
generalized equations, we must require that $v^A_B \equiv \delta^l g^A/\delta
a^B|_{a=0}$ locally has an inverse.
When this $v^{-1}$ exists, the generalized
Schwinger-Dyson equations are in a one-to-one correspondence with
those obtained from exploring invariance of the  $d\phi$-measure
(without the factor of $\rho(\phi)$) under local shifts. We will
assume that the space of fields forms a manifold. If the dimension of
the space is $N$ (i.e., $A = 1, \ldots , N$), we need precisely those symmetry
transformations that locally correspond to shifts. These
transformations are parametrized by $N$ fields $a^A(x)$.

The condition that the measure $d\phi\rho(\phi)$ be invariant under
the transformation (20) is equivalent to
\beq
\frac{\delta^r\rho}{\delta \phi^C} - (-1)^{\epsilon_A + \epsilon_C}
G^A_{CA}\rho = 0~,
\eeq
with
\beq
G^A_{CA} \equiv -\left(v^{-1}\right)^B_C\frac{\delta^r}{\delta
  \phi^A}\left(v^A_B\right) ~.
\eeq

How do we now find the modified Schwinger-Dyson BRST symmetry whose
Ward identities are the equations (21)? As in ref. \cite{us}, we can again
follow the collective
field approach. We do this by promoting $a^A(x)$ to a genuine
field in the path integral, which we integrate over by using a flat measure.
The relevant BRST symmetry reads \cite{us1}:
\begin{eqnarray}
\delta \phi'^A & = & - \left(M^{-1}\right)^A_B \frac{\delta^r g^B}{\delta
a^C} c^C \cr
\delta a^A & = & c^A \cr
\delta c^i & = & 0 \cr
\delta \phi^*_A & = & B_A \cr
\delta B_A & = & 0 ~,
\end{eqnarray}
where $M^A_B \equiv \delta^r g^A/\delta \phi'^B$. Nilpotency of the
transformations (24) is not immediately evident, but can be checked to hold:
$\delta^2 = 0$. This is also obvious from its construction in ref.
\cite{us1}. Next, we choose to gauge-fix on the trivial surface $a^A = 0$.
We do this by adding $-\delta[\phi^*_Aa^A] = (-1)^{\epsilon_A+1}B_Aa^A
- \phi^*_Ac^A$ to the
action $S$. At this point we can integrate out $B_A$ and $a^A$, modifying
the BRST transformations accordingly. The result is, for the BRST algebra:
\begin{eqnarray}
\delta \phi^A & = & -(-1)^{\epsilon_B(\epsilon_A+1)} v^A_Bc^B \cr
\delta c^A & = & 0 \cr
\delta \phi^*_A & = & (-1)^{\epsilon_B(\epsilon_A+1)}\frac{\delta^l S}
{\delta \phi^B}v^B_A ~,
\end{eqnarray}
where we have used the boundary condition $g^A(\phi',a\!=\!0) =
\phi'^A$.

One can readily check that the BRST Ward identities $0 = \langle \delta
\{\phi^*_A F[\phi]\}\rangle$ precisely coincide with the Schwinger-Dyson
equations (21). Equation (25) thus gives us the required Schwinger-Dyson BRST
algebra. However, nilpotency of the BRST operator is lost in the
process of integrating out $B_A$ and $a^A$. In contrast to the usual
case of $\rho = 1$ \cite{us}, nilpotency does not even hold in general
on the space of fields $\phi$ only. This makes this form of the
Schwinger-Dyson BRST algebra slightly awkward for the quantization
programme. But the version of the collective field formalism we have adhered
to until now corresponds to the ``Abelianization'' of the constraints.
As it turns out, the problem of nilpotency of the operator $\delta$
is instantly solved if we instead use the non-Abelian formalism (see
appendix A of ref. \cite{us1}). We shall now describe this in some
detail.\footnote{The notation follows Appendix A of ref. \cite{us1}.}
The non-Abelian Schwinger-Dyson BRST transformations can be chosen in
the form
\begin{eqnarray}
\delta \phi'^A &=& u^A_B(\phi')c^B \cr
\delta a^A &=& - \nu^A_B(a)c^B \cr
\delta c^A &=& - \frac{1}{2}(-1)^{\epsilon_B}c^A_{BC}c^Cc^B \cr
\delta \phi^*_A &=& B_A \cr
\delta B_A &=& 0~,
\end{eqnarray}
where
\beq
u^A_B(\phi') ~\equiv~
\left.\frac{\delta^r g^A(\phi',a)}{\delta a^B}\right|_{a=0}~.
\eeq
The supernumbers $c^A_{BC}$ are the structure coefficients of the
supergroup of transformations (20). They satisfy
\beq
c^A_{BC} ~=~ -(-1)^{\epsilon_B\epsilon_C}c^A_{CB} ~.
\eeq
A boundary condition is $\nu^A_B(a\!\!=\!\!0) = \delta^A_B$, and we also have
$\lambda^A_B(a)\nu^B_C(a) = \delta^A_C$ \cite{Hamermesh}.
We integrate the collective field
over the left- or right-invariant measure of the supergroup of
transformations $g^A$. The full
functional measure is then formally (i.e. with a symmetry-preserving
regulator) invariant under the BRST transformation (26) if we take, for $[dc]$
and $[d\phi^*]$, the usual (flat) measures, and if we assume
that the group of transformations is compact (and in particular
$(-1)^{\epsilon_A}c^A_{AB} = 0$). We now gauge-fix the collective
field $a^A$ to zero
by adding a term $-\delta[\phi^*_Aa^A] = (-1)^{\epsilon_A+1}B_Aa^A +
\phi^*_A\nu^A_B(a)c^B$
to the action $S$. Integrating over $B_A$ and $a_A$, we find the
modified BRST transformations by substituting for $B_A$ the equation
of motion for $a^A$ (at $a^A = 0$). It is important to take into account
the contribution from the measure as well. If we define
\beq
\bar{\Gamma}^A_{BC} \equiv \left.\frac{\delta^r\nu^A_B}{\delta
  a^C}\right|_{a=0} ~,
\eeq
then the BRST transformations can be written
\begin{eqnarray}
\delta \phi^A &=& u^A_B(\phi) c^B \cr
\delta c^A &=& -\frac{1}{2}(-1)^{\epsilon_B}c^A_{BC}c^Cc^B \cr
\delta \phi^*_i &=& (-1)^{\epsilon_A}\frac{\delta^l S}{\delta\phi^B}
u^B_A(\phi) + i\hbar(-1)^{\epsilon_A+\epsilon_B}\bar{\Gamma}^B_{BA} +
(-1)^{\epsilon_A\epsilon_B}\phi^*_M\bar{\Gamma}^M_{BA}c^B ~.
\end{eqnarray}
A related BRST construction for field
theories with vanishing equations of motion, $\delta
S/\delta\phi^A = 0$ has been considered by Okubo \cite{Okubo}. One has
$\bar{\Gamma}^G_{KL} - (-1)^{\epsilon_K\epsilon_L}\bar{\Gamma}^G_{LK}
= c^G_{KL}$.

Due to the ``quantum correction'' to the transformation law for
$\phi^*_A$, the action $S$ itself is not invariant under the
transformations (30). However, the measure transforms in just such a
manner as to cancel the remaining term. So the combination of action
and measure is invariant under (30), as it should be. The last
two terms in the transformation law for $\phi^*_A$ cancel when we
consider the Ward identity $0 =
\langle\delta[\phi^*_AF(\phi)]\rangle$, leaving us with the correct
Schwinger-Dyson equations.

We now perform the change of variables
\beq
C^A = u^A_B(\phi)c^B~,~~~~~ \Phi^*_A =
\phi^*_B\left(u^{-1}\right)^B_A~.
\eeq
The result is:
\begin{eqnarray}
\delta\phi^A &=& C^A \cr
\delta C^A &=& 0 \cr
\delta\Phi^*_A &=& \frac{\delta^l S}{\delta \phi^A} +
(-1)^{\epsilon_M+1}\Gamma^M_{BA}C^B\Phi^*_C
+ i\hbar(-1)^{\epsilon_A+\epsilon_C}\bar{\Gamma}^C_{CB}
\left(u^{-1}\right)^B_A ~,
\end{eqnarray}
where $\Gamma^A_{BC}$ is defined to be
\beq
\Gamma^A_{BC} = G^A_{BC} +
(-1)^{\epsilon_A(\epsilon_M+\epsilon_C+1)}u^M_S\bar{\Gamma}^S_{CB}
\left(u^{-1}\right)^B_A\left(u^{-1}\right)^C_K~,
\eeq
and where we have introduced the connection\footnote{This definition is
consistent with the one given in eq. (23).}
\beq
G^D_{AC}(\phi) ~=~ (-1)^{\epsilon_A(\epsilon_D+1)}u^D_B
\frac{\delta^r \left(u^{-1}\right)^B_A}{\delta\phi^C}~.
\eeq

The action $S$ is again not invariant under the BRST transformation,
but the full partition function is,
provided that $\rho$ is covariantly conserved with respect to
$G^A_{BC}$:
\beq
\frac{\delta\rho}{\delta\phi^A} - (-1)^{\epsilon_A+\epsilon_B}\rho(\phi)
G^B_{AB} = 0 ~.
\eeq
But this is just the condition (22) that the measure
$d\phi\rho(\phi)$ is invariant under the group of transformation
$g^A$. So we again find that the combination of action and measure is
invariant under this (now non-Abelian) Schwinger-Dyson BRST
transformation.

The advantage of this non-Abelian formulation is that nilpotency of
$\delta$ when acting on the space of fields $\phi^A$ is not lost in
the process of integrating out the collective field $a^A$ and the
Nakanishi-Lautrup field $B_A$. This means that the BRST operator
$\delta$ can be used to gauge-fix internal gauge symmetries as well,
and it is therefore meaningful to formulate the quantization prescription
in terms of a Lagrangian Master Equation. This equation follows again
from the simple requirement that the combination of action and measure
remain invariant under the Schwinger-Dyson BRST symmetry. Let us write
$S_{ext} = S^{BV}[\phi,\Phi^*] + \Phi^*_AC^A$. Since $S^{BV}$ now depends
on $\Phi^*$, we find again that the transformation
law for $\Phi^*_A$ has to be modified slightly. The resulting
transformation is
\beq
\delta\Phi^*_A = \frac{\delta^l S^{BV}}{\delta\phi^A} + (-1)^{\epsilon_M +1}
\Gamma^M_{AK}C^K\Phi^*_M + (-1)^{\epsilon_A\epsilon_M}\frac{\delta^r
S^{BV}}{\delta\Phi^*_B}\Gamma^M_{BA}\Phi^*_M + i\hbar(-1)^{\epsilon_A +
\epsilon_C}\bar{\Gamma}^C_{CB}\left(u^{-1}\right)^B_A ~,
\eeq
with the transformations for $\phi$ and $C$ left untouched. Note that
only $S^{BV}$ enters in the transformation law for $\Phi^*$. The condition
that the path integral remains Schwinger-Dyson BRST-invariant
leads precisely
to the standard Master Equation for $S^{BV}$:
\beq
\frac{\delta^r S^{BV}}{\delta\Phi^*_A}\frac{\delta^l S^{BV}}{
\delta\phi^A} = -i\hbar \Delta_\rho S^{BV} ~.
\eeq
We wish to emphasize that in the present formulation this is a highly
non-trivial result of delicate cancellations between action and measure,
as well as of the continuity equation (35).

Boundary conditions need to be imposed on $S^{BV}$. A first requirement
is that $S_{cl}[\phi] = S^{BV}[\phi,\Phi^*\!=\!0]$,
where $S_{cl}$ is the
classical action. This is needed to ensure that Schwinger-Dyson equations
for $S_{ext} = S^{BV}[\phi,\Phi^*] + \Phi^*_AC^A$ formally agree with
those of $S_{cl}$ before any of
the possible internal gauge symmetries have been fixed.\footnote{For the
special case of no internal gauge symmetries, $S^{BV}[\phi,\Phi^*]
= S_{cl}[\phi]$, and it is then straightforward to see that the Ward
identities of the symmetry (32) and (36) yield the correct
Schwinger-Dyson equations for $S_{cl}[\phi]$.} One further
boundary condition is needed to ensure regularity of $S^{BV}$, i.e.,
invertibility of the propagator matrix.

We see that knowing the group of transformations that leave the
measure $d\phi\rho(\phi)$ invariant naturally leads to an object
($G^A_{BC}$) that transforms as a connection on the space
of fields. This connection itself has only indirect physical significance,
since just the traced-over object $(-1)^{\epsilon_A}G^A_{CA}$ appears
in the Schwinger-Dyson equations.
Note that the Schwinger-Dyson BRST transformations (32) are ambiguous
as far as
the connection is concerned. We can replace any suitable connection
$G^A_{BC}$
with $G^A_{BC} + \tilde{G}^A_{BC}$ as long as $\tilde{G}^A_{BC}$ has
the correct symmetry properties under exchange of the lower indices, and as
long as $(-1)^{\epsilon_A}\tilde{G}^A_{BA} = 0$. Such a replacement is
void of physical content. It is conceivable that the redundancy in the
choice of connection is a reflection of the large group of symmetries
of the covariant Master Equation. If so, this could permit a geometric
interpretation of the group of invariances directly on the space of fields.

We end this paper with some general comments.
We have throughout restricted ourselves to either transformations of
the fields $\phi^A$ that do not depend on the ghosts $c^A$ or
antighosts (``antifields'' in the language of Batalin and Vilkovisky)
$\phi^*_A$, or, in the last part, on symmetries of functional measures
of the fields $\phi^A$ only. We have done this on the assumption that
eventually only symmetry properties related to the original classical
fields (part of $\phi^A$) are of physical importance. This means that
we have really only been interested in the subset of transformations
involving $\phi^A$ that refer to the classical fields, and not to the
usual ghosts, antighosts, auxiliary fields, ghosts-for-ghosts, etc.,
which may be required to complete the quantization programme, and
which form another part of $\phi^A$. Such a
point of view may be too restrictive, and there is indeed nothing
preventing a more general setting in which all fields $\phi^A$ are mixed
with each other and
with ghosts $c^A$ and antighosts $\phi^*_A$. These more general
transformations must of course obey the quite restrictive condition of
preserving Grassmann parities and ghost numbers. The discussion in
refs. \cite{Geometry,Schwarz} goes along such lines (for
the case where the ghosts $c^A$ have been
integrated out, and where the remaining fields $\phi^A$ and antighosts
$\phi^*_A$ thus are canonical variables under the antibracket). One may
in that case phrase the canonical framework in terms of a supersymplectic
formalism that resembles the usual symplectic
formulation of classical Hamiltonian mechanics. The different ghost
number and Grassmann parity assignments between ``coordinates''
($\phi^A$) and ``momenta'' ($\phi^*_A$) does, however, make the
analogy with classical mechanics somewhat limited.
It is difficult and rather tedious to formulate correct
boundary conditions to be imposed on the Master Equations in any other
frame than that of (the analogue of) Darboux coordinates on the
supersymplectic manifold.

As we have shown in this letter,
the analogue of Batalin-Vilkovisky quantization on spaces with
non-trivial measure densities can be derived straightforwardly from
the underlying Schwinger-Dyson BRST algebra. It is not coincidental
that upon
integrating out the ghosts $c^A$, the Master Equation for theories
with non-trivial $\rho(\phi)$-measures formally matches
the one of Schwarz \cite{Schwarz}, although we have not made use of the
fact that a Darboux frame exists in which $\rho = 1$. This
is because the existence of such a frame is a sufficient
but not necessary condition for having nilpotency of the operator
$\Delta_\rho$. As we have seen, the existence of a
coordinate frame with, in the language of ref. \cite{Schwarz},
$\rho(\phi,\phi^*) = \rho(\phi)$, also ensures that $\Delta_{\rho}^2 = 0$.
It is only when
leaving this density $\rho(\phi)$ in the measure (instead of exponentiating
it into a ``one-loop correction'' of the extended action) that the
full geometric picture discussed in this paper emerges.

\vspace{0.5cm}

\noindent
{\sc Acknowledgement:} ~We wish to thank M. Henneaux for interesting
discussions. The work of J.A. was supported in part by Fondecyt 1930566.

\end{document}